\documentclass{ws-procs9x6}            
\usepackage{subfigure}
\def\Tr{{\rm Tr}}
\newcommand{\sla}{\not\!}
\begin{document}
\title{Interactions between two heavy mesons within heavy meson chiral effective field theory}

\author{Zhan-Wei Liu$^{1,2,*}$, Bo Wang$^{3}$, Hao Xu$^{4}$, Ning Li$^{5}$, Xiang Liu$^{1,2}$, Shi-Lin Zhu$^{3,6,7}$}

\address{$^1$School of Physical Science and Technology, Lanzhou University, Lanzhou 730000, China\\
$^2$Research Center for Hadron and CSR Physics,
Lanzhou University $\&$ Institute of Modern Physics of CAS,
Lanzhou 730000, China\\
$^3$Center of High Energy Physics, Peking University, Beijing 100871, China\\
$^4$Department of Applied Physics, School of Science, Northwestern  Polytechnical University, Xi'an 710072, China\\
$^5$Facility for Rare Isotope Beams and Department of Physics and Astronomy,
Michigan State University, East Lansing, MI 48824, USA\\
$^6$School of Physics and State Key Laboratory of Nuclear Physics and Technology, Peking University, Beijing 100871, China\\
$^7$Collaborative Innovation Center of Quantum Matter, Beijing 100871, China\\
$^*$E-mail: liuzhanwei@lzu.edu.cn
}

\begin{abstract}
We have studied the interactions between two heavy mesons ($D^{(*)}$-$D^{(*)}$, $\bar D^{(*)}$-$\bar D^{(*)}$, $B^{(*)}$-$B^{(*)}$, or $\bar B^{(*)}$-$\bar B^{(*)}$) within heavy meson chiral effective field theory and investigated possible molecular states. The effective potentials are obtained with Weinberg's scheme up to one-loop level. At the leading order, four body contact interactions and one pion exchange contributions are considered. In addition to two pion exchange diagrams, we include the one-loop chiral corrections to contact terms and one pion exchange diagrams at the next-to-leading order. The effective potentials both in momentum space and coordinate space are investigated and discussed extensively. The possible molecular states are also studied and the binding energies are provided by solving the Schrodinger equation. The results will be helpful for the experimental search for the doubly-heavy  molecular states.
\end{abstract}


\bodymatter

\section{Introduction}\label{aba:sec1}
Hadron structures have still not been clearly disclosed since the first hadron, proton, was discovered in the beginning of last century. Although the quark model can deal with most hadron phenomenology, more and more exotic behaviors have been noticed in hadron physics in experiment. The structures are closely related to the hadron interactions at low energies, and the relevant studies would also help us to understand the non-pertubative properties of QCD.

The low-lying baryons, $N(1440)$, $N(1535)$, and $\Lambda(1405)$, cannot be interpreted well even only for their masses within conventional quark model. After considering the hadron interactions between $\pi N$, $\bar K N$, and so on, we can obtain the reasonable mass poles for them within Hamiltonian effective field theory, and moreover, both the scattering data at experiment and the lattice QCD simulations can studied well with such a scheme \cite{Liu:2015ktc,Liu:2016wxq,Liu:2016uzk,Wu:2017qve}. In the heavy flavor sector, the low-mass puzzle for $\Lambda_c(2940)$ can be resolved with the interaction of $D^* N$ included \cite{Luo:2019qkm}.

The exotic hadrons totally without the bare core of conventional quark model are more sensitive to the hadron interactions. Recently discovered $Z_b$ and $P_c$ states are probably such molecular states \cite{Collaboration:2011gja,Aaij:2019vzc,Aaij:2015tga}. $B^{(*)}$ and $\bar B^{(*)}$ are attracted by each other and can form hidden-bottom bound states, probably $Z_b(10610)$ and $Z_b(10650)$. In this proceeding, we focus on doubly-heavy molecules.

In this proceeding, we systematically investigate the interaction between two heavy mesons, $D^{(*)}$-$D^{(*)}$, $\bar D^{(*)}$-$\bar D^{(*)}$, $B^{(*)}$-$B^{(*)}$, or $\bar B^{(*)}$-$\bar B^{(*)}$, within chiral effective theory. After deriving the effective potentials for the two interacting heavy mesons in Sec. \ref{sec.effP}, we solve the Schr\"odinger equation to study the doubly-heavy molecular states in Sec. \ref{sec.mol}. Since doubly-charmed baryon $\Xi_{cc}^{++}$ has been observed at LHCb \cite{Aaij:2017ueg}, we believe such doubly-heavy molecules could be detected in experiment soon.

\section{Interactions between two heavy mesons} \label{sec.effP}
Chiral effective field theory can systematically deal with the low-energy processes order by order. The results are expanded in power series of small momenta, small masses or residual masses, and so on, rather than the strong coupling $\alpha_s$. At first we need to construct the Lagrangians up to the next-to-leading order. The details can be found in Refs. \cite{Liu:2012vd,Wang:2018atz,Xu:2017tsr}. Here we only list the Lagrangians at leading order.
\begin{eqnarray}
\mathcal L^{(0)}_{4H}&=&D_{a}\Tr[H \gamma_\mu \bar H ]\Tr[ H
\gamma^\mu\bar H] +D_{b}\Tr[H \gamma_\mu\gamma_5 \bar H ]\Tr[ H
\gamma^\mu\gamma_5\bar H] \nonumber\\&& 
\hspace{-2em}+E_{a} \Tr[H
\gamma_\mu\lambda^a \bar H ]\Tr[ H \gamma^\mu\lambda_a\bar H]
+E_{b}\Tr[H \gamma_\mu\gamma_5\lambda^a \bar H ]\Tr[ H
\gamma^\mu\gamma_5\lambda_a\bar H],  \label{HHHH}
\\
\mathcal L^{(1)}_{H\phi}&=&-\langle (i v\cdot \partial H)\bar H
\rangle
                         +\langle H v\cdot \Gamma \bar H \rangle
                         +g\langle H \sla u \gamma_5 \bar H\rangle
                         -\frac18 \Delta \langle H \sigma^{\mu\nu} \bar H \sigma_{\mu\nu} \rangle.\label{L1}
\end{eqnarray}
Here $H$ and $\bar H$ are related to the $B^{(*)}$ field, $\Gamma$ and $u$ contain the pseudoscalar mesons. The first Lagrangian in Eq. (\ref{HHHH}) can describe the contact interactions $\bar B^{(*)}$-$\bar B^{(*)}$-$\bar B^{(*)}$-$\bar B^{(*)}$, and the second in Eq. (\ref{L1}) contains the vertice like $\bar B^{(*)}$-$\bar B^{(*)}$-$\pi$, $\bar B^{(*)}$-$\bar B^{(*)}$-$\pi$-$\pi$.

\begin{figure}[!htbp]
\centering
\scalebox{1}{\includegraphics{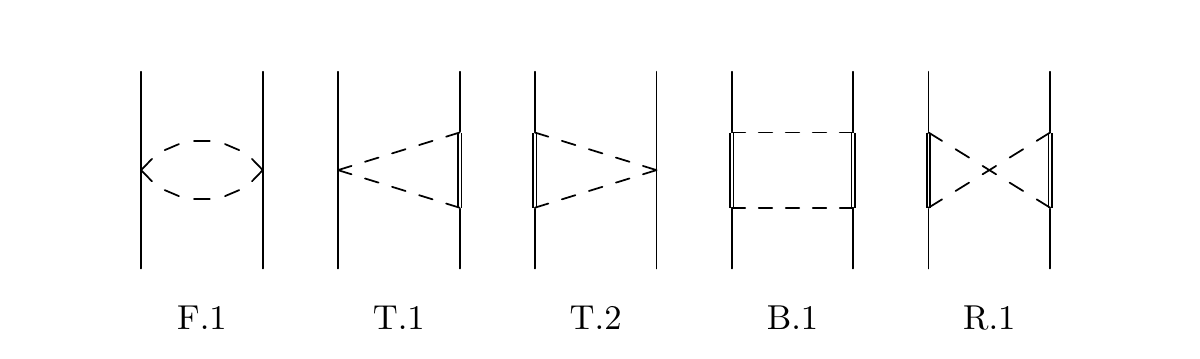}}\\
\caption{The 2$\pi$-exchange diagrams including the football
diagram (F.1), triangle diagrams (T.1 and T.2), box diagram (B.1),
and crossed diagram (R.1).
 } \label{LoopTPi}
 \end{figure}

After constructing the Lagrangians, we can derive the effective potentials in momentum space from calculating Feynman diagrams in Weinberg scheme \cite{Weinberg:1990rz}. At leading order, there are contact and one-pion-exchange diagrams in which the vertice are from the Lagrangians at leading order. At next to leading order, there are four types of diagrams. The first is from two-pion-exchange diagrams shown in Fig. \ref{LoopTPi}. The second and third are loop corrections to the contact terms and the one-pion-exchange terms. The fourth is from the tree diagrams with vertice in the Lagrangians at the next-to-leading order, which can absorb the ultraviolet divergences from the loop diagrams.

We calculate these diagrams using dimensional regularization and modified minimal subtraction scheme. Before providing the numerical potentials, there are still some unknown low-energy constants which cannot be determined with chiral effective field theory itself. We try to estimate them via combining the resonance saturation method, SU(4) approximation, non-analytic approximation, fitting the lattice QCD results, and so on.

\begin{figure}[!htbp]
 \centering
 \subfigure[$I=0$]{
\scalebox{0.26}{\includegraphics{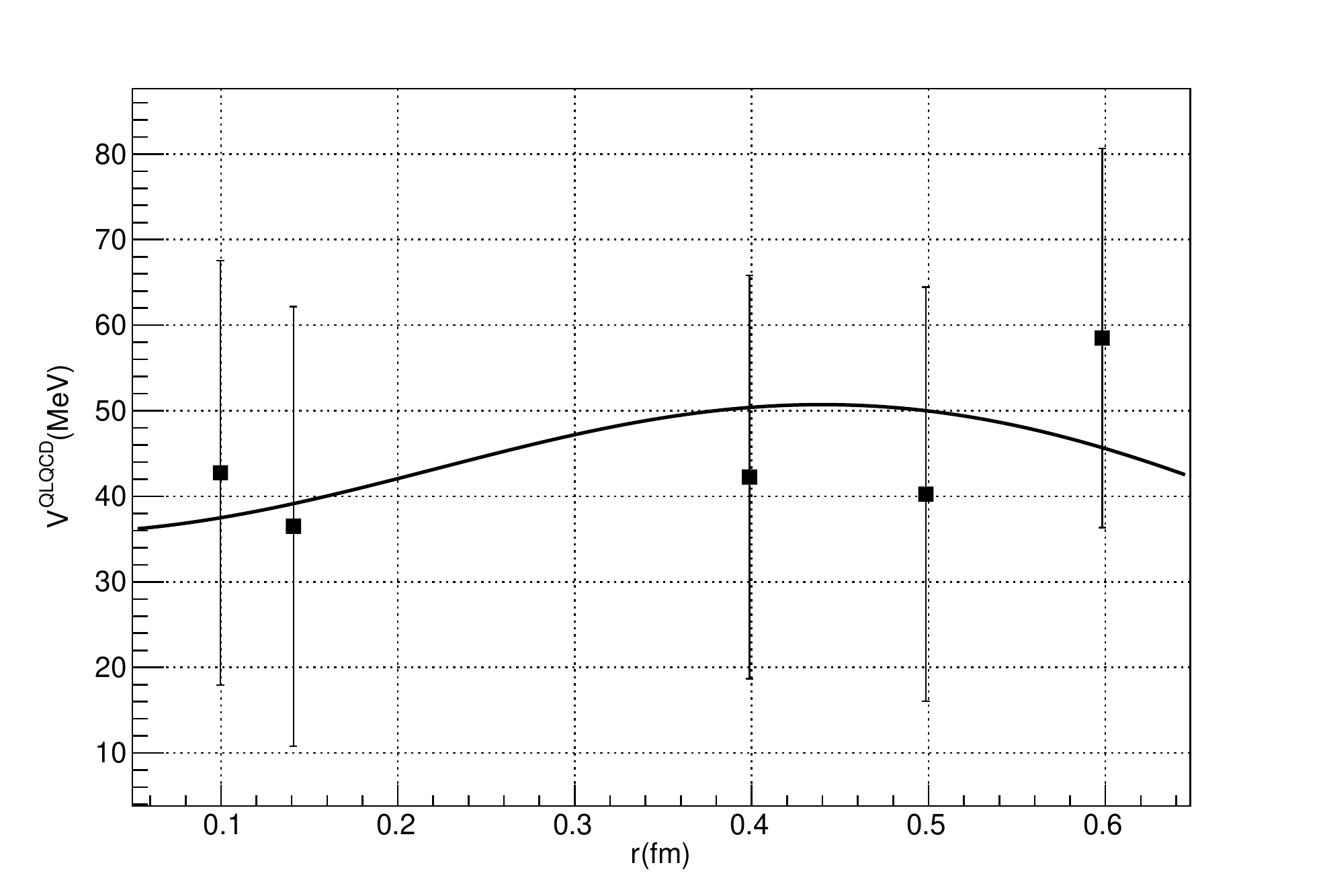} }
 }
 \subfigure[$I=1$]{
\scalebox{0.26}{\includegraphics{zisoZr8a} }
 }
 \caption{Fitting the $\bar B\bar B$ potentials.
 The data of the quenched lattice QCD are derived from Ref. \cite{Detmold:2007wk} where $m_\pi=402.5\pm6.7$ MeV.}
 \label{figFit}
 \end{figure}

\begin{figure}[!htbp]
 \centering
 \subfigure[momentum space]{
\scalebox{0.17}{\includegraphics{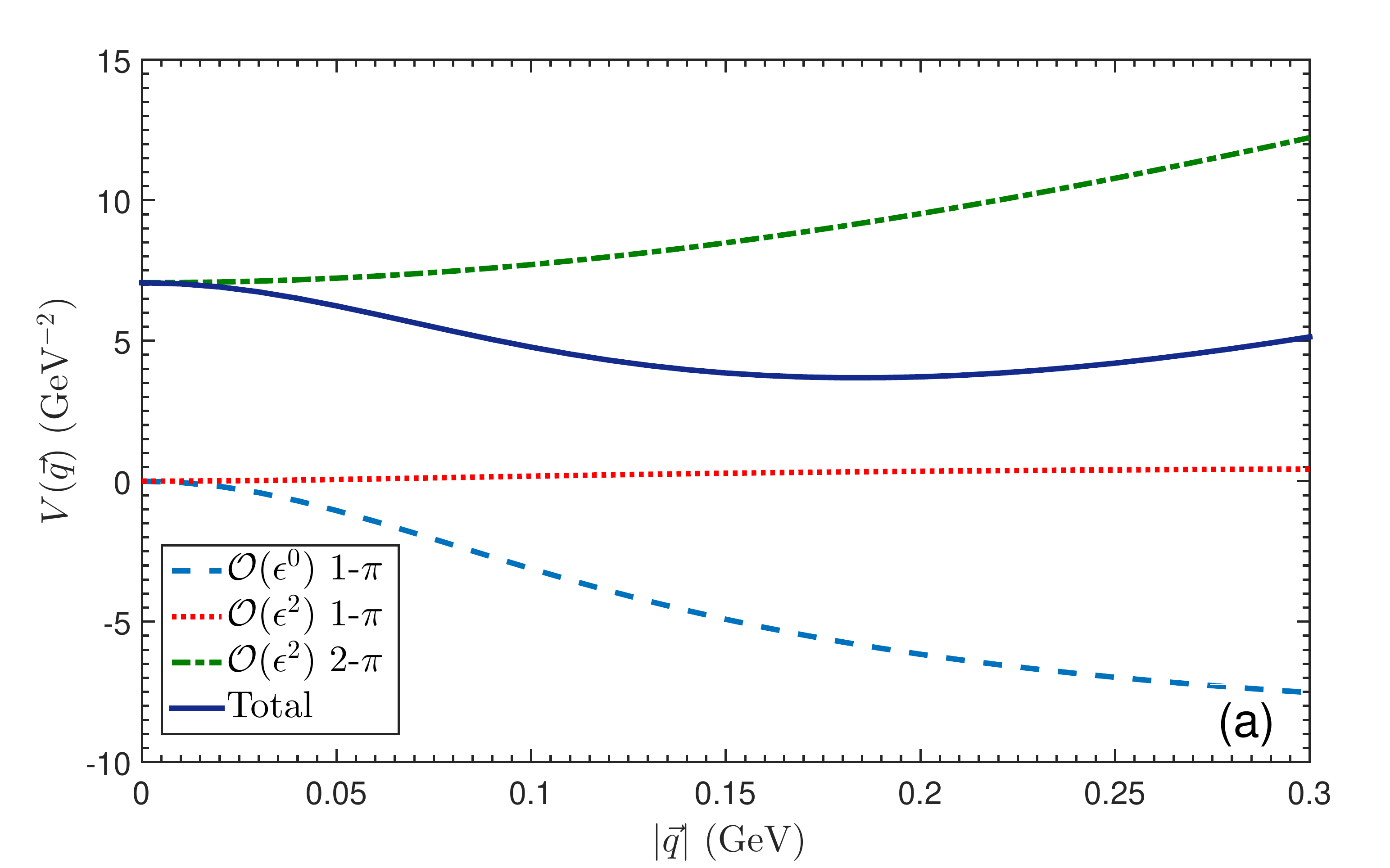}}
 }
 \subfigure[coordinate space]{
\scalebox{0.17}{\includegraphics{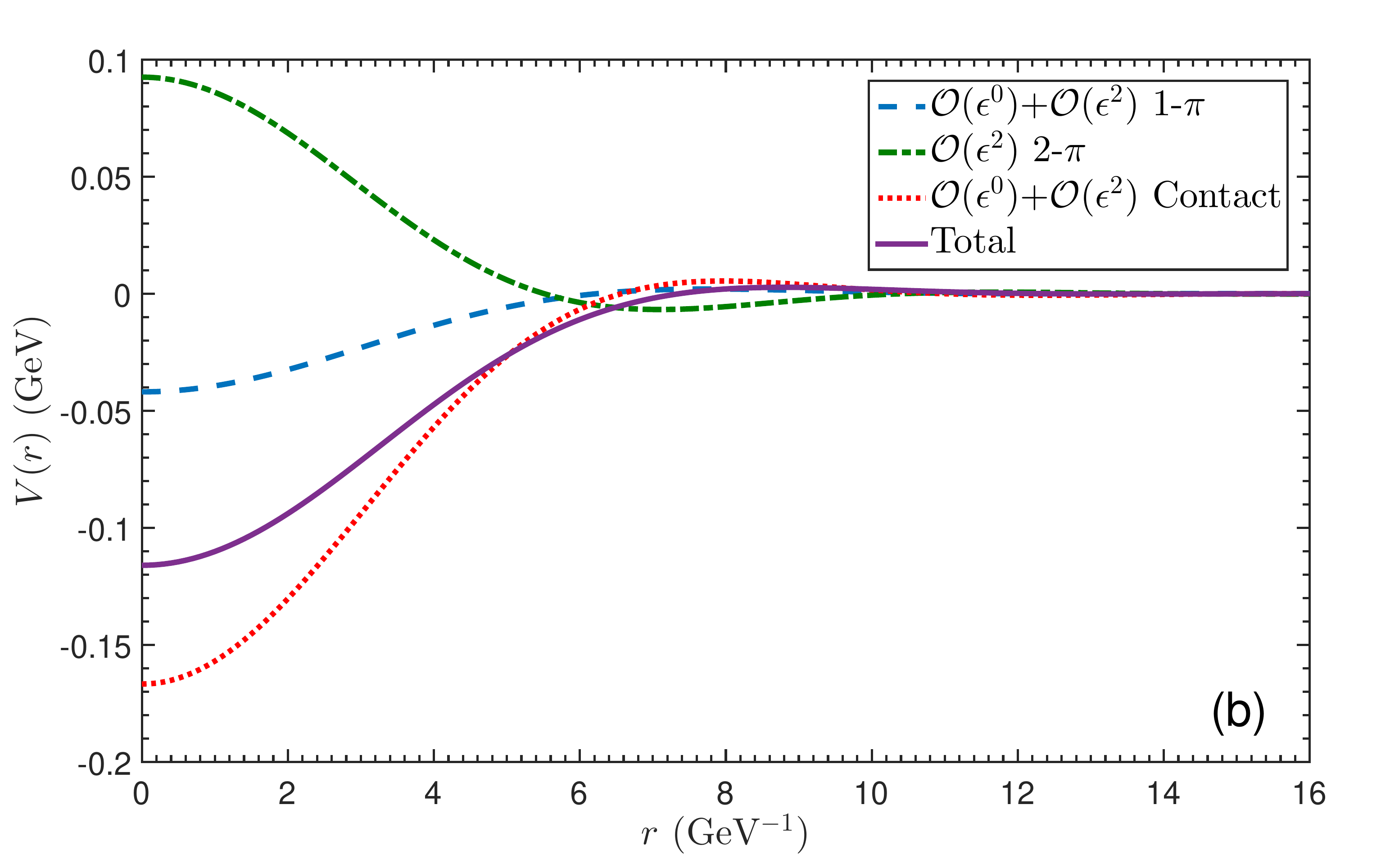}}
 }
 \caption{Effective potentials for the $\bar B^*\bar B^*$ with $I(J^P)=0(1^+)$}
 \label{figPtl}
 \end{figure}

Now we can obtain the effective potentials in momentum space. We show the potentials for the $\bar B^*\bar B^*$ with $I(J^P)=0(1^+)$ in the left diagram of Fig. \ref{figPtl} as an example. From the figure, one notices that the corrections at the next-to-leading order cannot be neglected.

\section{Doubly-heavy molecular states} \label{sec.mol}

With the effective potentials obtained within chiral effective field theory, we can investigate the doubly-heavy molecular states made of $\bar B^{(*)}$-$\bar B^{(*)}$ or $D^{(*)}$-$D^{(*)}$. First, we need to get the effective potentials in coordinate space from those in momentum space in Sec. \ref{sec.effP} via Fourier transformation. As an example, we present the potentials in coordinate space for the $\bar B^*\bar B^*$ with $I(J^P)=0(1^+)$ in the right diagram of Fig. \ref{figPtl}. One can study the properties of the molecular states via solving the Schr\"odinger equation in coordinate space.

The possible molecular states are listed in Refs. \cite{Wang:2018atz,Xu:2017tsr}. Among them, there are two bound states in the channels of $\bar{B}\bar{B}^\ast$ and $\bar{B}^\ast\bar{B}^\ast$ with $I(J^P)=0(1^+)$ with the masses being $m_{\bar{B}\bar{B}^\ast}\simeq10591.4^{+9.2}_{-12.9}$ MeV, $m_{\bar{B}^\ast\bar{B}^\ast}\simeq10625.5^{+16.3}_{-21.5}$ MeV. They are below the threshold of $\bar B\bar B\pi$, and thus the strong decay is forbidden. These two molecules are stable under the strong interactions and can be searched for in the electromagnetic channels such as $\bar B\bar B\gamma$,  $\bar B\bar B\gamma\gamma$, and so on.

This project is supported by the National Natural Science Foundation of China under Grants No. 11705072 and 11965016, 11575008, 11621131001, and the China National Funds for Distinguished Young Scientists under Grants No. 11825503.


\bibliographystyle{ws-procs9x6} 


\end{document}